\documentclass[runningheads]{apinv}  %citeauthoryear
\usepackage{epsfig,cite,graphics}
\usepackage{marvosym} %For astronomical symbols % % %
\usepackage[utf8]{inputenc}
\usepackage{natbib}
\newcommand{\hvezda}{KIC\,6950556}

\bibpunct{(}{)}{;}{a}{}{,}
\begin{document}

\title{Fine detrending of raw Kepler and MOST \\ photometric data of KIC 6950556 and HD 37633}
\titlerunning{Fine detrending of raw Kepler and MOST photometric data}
\author{Zden\v{e}k Mikul\'{a}\v{s}ek\inst{1}, Ernst Paunzen\inst{1}, Miloslav Zejda\inst{1}, Evgenij Semenko\inst{2},\\ Klaus Bernhard\inst{3}, Stefan H\"{u}mmerich\inst{3}, Jia Zhang\inst{1,4}, Swetlana Hubrig\inst{5}, Rainer~Kuschnig\inst{6,7}, Jan Jan\'{i}k\inst{1}, Miroslav Jagelka\inst{1}}
\authorrunning{Z. Mikul\'{a}\v{s}ek, E. Paunzen, M. Zejda et al.}
\tocauthor{Zden\v{e}k Mikul\'{a}\v{s}ek, Ernst Paunzen, Miloslav Zejda, Evgenij Semenko, Klaus Bernhard, Stefan H\"{u}mmerich, Swetlana Hubrig, Rainer Kuschnig, Jia Zhang, Jan Jan\'{i}k, Miroslav Jagelka}
% Command tocautor{} is used by the Latex to give author names
% to the Contents of the volume (automatically generated)
\institute{Department of Theoretical Physics and Astrophysics, Masaryk University,\\ Kotl\'{a}\v{r}sk\'{a} 2, cz\,611\,37 Brno, Czech Republic
	\and Special Astrophysical Observatory of the RAS, Russia
	\and Bundesdeutsche Arbeitsgemeinschaft für Ver\"{a}nderliche Sterne e.V. (BAV),\\ Berlin, Germany
    \and National Astronomical Observatories/Yunnan Observatory, Chinese Academy of Sciences, Kunming, China
    \and Leibniz-Institut für Astrophysik Potsdam (AIP), Potsdam, Germany
    \and University of Vienna, Institute for Astronomy, Vienna, Austria
    \and Department of Physics and Astronomy, University of British Columbia,\\Vancouver, Canada
    \newline
	\email{mikulas@physics.muni.cz}    }
\papertype{Submitted on xx.xx.xxxx; Accepted on xx.xx.xxxx}	
% Papertype can be "Research report", "Review", "Invited lecture", "Conference talk",
% "Conference poster", "Lecture at scientific seminar", "Summary of dissertation",  etc.
\maketitle

\begin{abstract}
We present a simple phenomenological method for detrending of raw Kepler and MOST photometry, which is illustrated by means of photometric data processing of two periodically variable chemically peculiar stars, KIC 6950556 and HD 37633. In principle, this method may be applied to any type of periodically variable objects and satellite or ground based photometries. As a by product, we have identified KIC 6950556 as a magnetic chemically peculiar star with an ACV type variability.
\end{abstract}

\section*{Introduction}
Nowadays, the amazing accuracy of nearly continuous photometric observations made from space satellites enables us to study subtle details in the light variations of variable stars. However, raw space photometric data contain different systematic trends associated with the spacecraft, detector and environment rather than the target. To exploit maximum information hidden in satellite photometry, we need to eliminate all these trends as good as possible. We have developed a way of detrending raw photometry using appropriate phenomenological models of the real light curve modulated by instrumental effects. To display the efficiency of the method, we have processed raw photometric data of two strictly periodic CP stars with smooth light curves obtained by the Kepler and MOST satellites.

\section*{1. Kepler – \hvezda}

\begin{figure}[!htb]
  \begin{center}
    \includegraphics[angle=90,width=1.00\textwidth,clip]{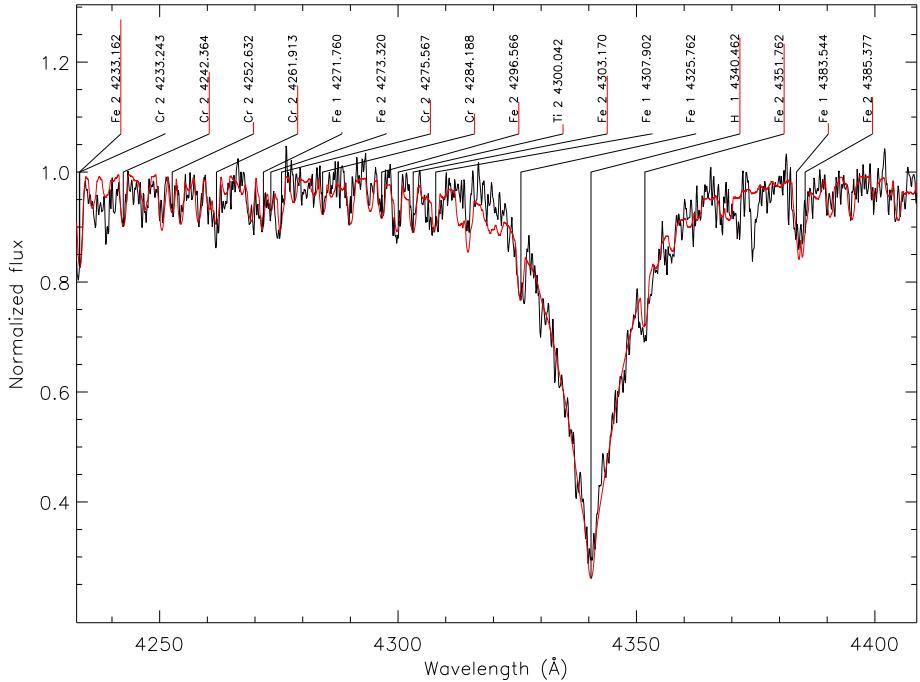}
    \caption[]{A part of the spectrum of SiCr CP2 star \hvezda, fitted by the atmosphere model with an effective temperature of 11\,000 K, $\log g=4.0$, and enhanced abundances of chromium, iron, and silicon.}
    \label{countryshape}
  \end{center}
\end{figure}
\hvezda\ (= 2MASS J19294376+4229306; $12.751 \pm 0.024$ mag) was classified by  \citet{mcnamara} as a B-type binary or a star whose light curve is dominated by rotation. Furthermore, it was preliminarily identified as an ellipsoidal (ELL) variable star candidate with a period of 1.5116\,d  by \citet{balon}. According to the double-wave light changes and the effective temperature, we also assumed ACV ($\alpha_2$\,CVn) variability. The classification of \hvezda\ as a magnetic chemically peculiar star was subsequently confirmed by our analysis of two newly obtained spectra in the spectropolarimetric mode using the 6m telescope of Special Astrophysical Observatory of the Russian Academy of Science, Nizhnii Arkhyz (see~Fig.\,\ref{countryshape}).

\begin{figure}[!htb]
  \begin{center}
    \includegraphics[width=0.85\textwidth,clip]{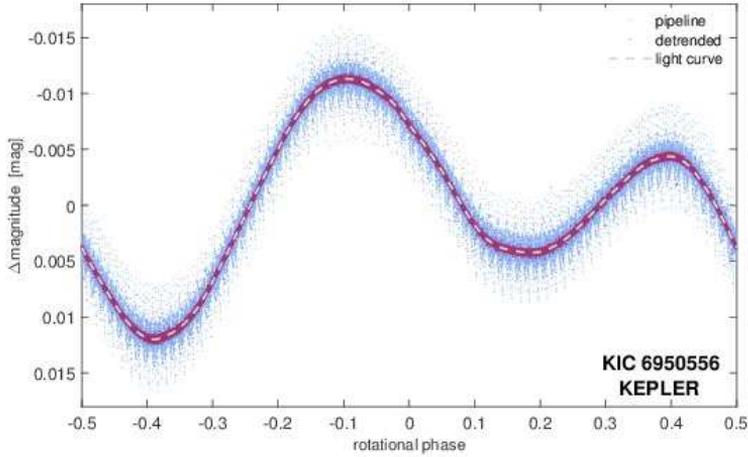}
    \caption[]{Phased light curve of \hvezda, based on photometric data detrended by the Kepler pipeline (powder-blue dots) and data detrended by our code (magenta dots). The dashed grey line is a fit of the light variations by a harmonic polynomial of the 35-th order.}
    \label{kepler2}
  \end{center}
\end{figure}
Spectral analysis of spectrograms shows that \hvezda\ is B8-9 V star with an effective temperature of $T_{\mathrm{eff}}=11\,000(700)$\,K, $\log g=3.5-4.0$, and a projected rotational velocity $V\sin i=60(10)$\,km\,s$^{-1}$. Assuming a typical radius of the stars with such parameters $R=2.4$\,R$_\odot$ \citep[according to tables in][]{harm} we can also estimate the inclination angle $i\doteq50^{\circ}$. The observed spectrum is apparently peculiar due to strong overabundance of chromium, iron, silicon, and underabundance of helium and calcium in the stellar atmosphere. If overabundant elements are distributed unevenly on the surface of the rigidly rotating star, they should form photometrically contrasting spots causing light variability of the star.

\begin{figure}[!htb]
  \begin{center}
    \includegraphics[width=0.85\textwidth,clip]{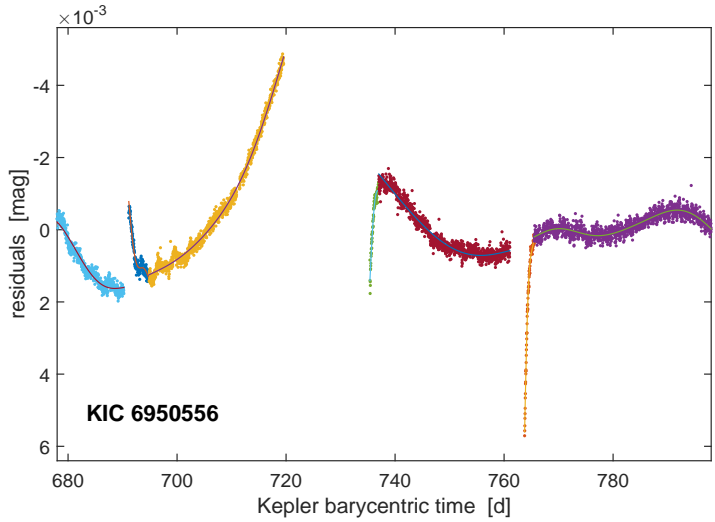}
    \caption[]{Detail view of seven residual segments fitted by special function series.}
    \label{residua4}
  \end{center}
\end{figure}

The photometric analysis was based exclusively on raw Kepler data (Q0-Q17) which exhibit a scatter of 32\,mmag as processed by the Kepler pipeline \citep[][]{jenkins}. Such a preprocessed set of 64\,793 individual observations of \hvezda\ exhibits a scatter of 1.0\,mmag – see Fig.\,\ref{kepler2}.
\begin{figure}[!htb]
  \begin{center}
    \includegraphics[width=1\textwidth,clip]{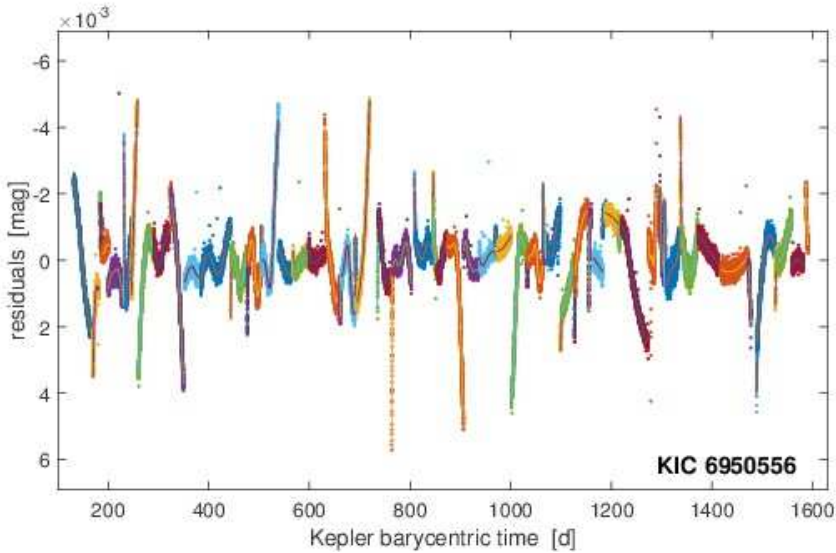}
    \caption[]{Residuals fitted by 75 smooth lines.}
    \label{residua3}
  \end{center}
\end{figure}
The detailed analysis of the light residuals indicates a very strong correlation of adjacent points (see Fig.\,\ref{residua4}) which can be removed by fine detrending. Residuals were divided into 75 intervals of uneven lengths and fitted by series of special quasi-orthogonal functions \citep{mik} of 0 to 7-th orders (see Fig.\,\ref{residua3}). After the detrending, which is described by 256 free parameters, we obtained a corrected light curve with a scatter of 0.13 mmag, which is close to the expected limit of accuracy of the Kepler satellite photometry. The detailed description of the mathematics of the detrending procedure will be published elsewhere.

The light curve itself was fitted by a harmonic polynomial of the 35-th order (see Fig.\,\ref{kepler2}). The accuracy ($5\times 10^{-7}$ mag) and reliability of the detrended light curve is sufficient for deriving the positions of photometric spots on the surface of the rotating star. The complex shape of the light curve with unequally high maxima in the double-wave \citep[see e.g.][]{morris} also do not support the classification of \hvezda\ as an elliptical variable star or B-type binary \citep[][]{balon,mcnamara}.

We also refined the period of the star to: $P=1.511\,785\,08(4)$ d, which seems to be fairly constant: $\dot{P}= 1(2) \times 10^{-10}$. It should be noted that the accuracy of the period rate determination of $2\times10^{-10}$ is high enough to reveal all presently known CP stars exhibiting variable periods \citep[][]{mikcp,mikbs}.

\section*{2. MOST – HD 37633}
The MOST B9Vp CP2 star HD\,37633 $(=V1147$ Ori) was identified as a variable star by \citet{north82}. Later, \citet{north84} published its period $P=1.5718$\,d and light curves of the star in the Geneva system.
\begin{figure}[!htb]
  \begin{center}
    \includegraphics[width=0.78\textwidth,clip]{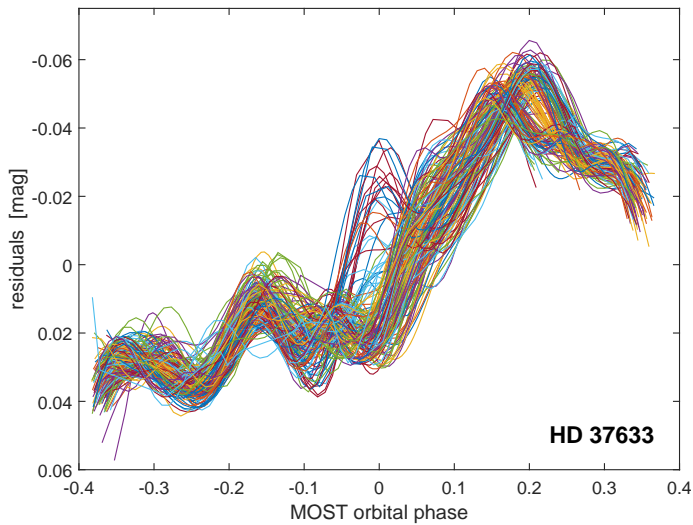}
    \caption[]{Residuals plotted versus the MOST orbital phase. Each of the 146 MOST passages were fitted by its own curve.}
    \label{most4}
  \end{center}
\end{figure}
\begin{figure}[!htb]
  \begin{center}
    \includegraphics[width=0.78\textwidth,clip]{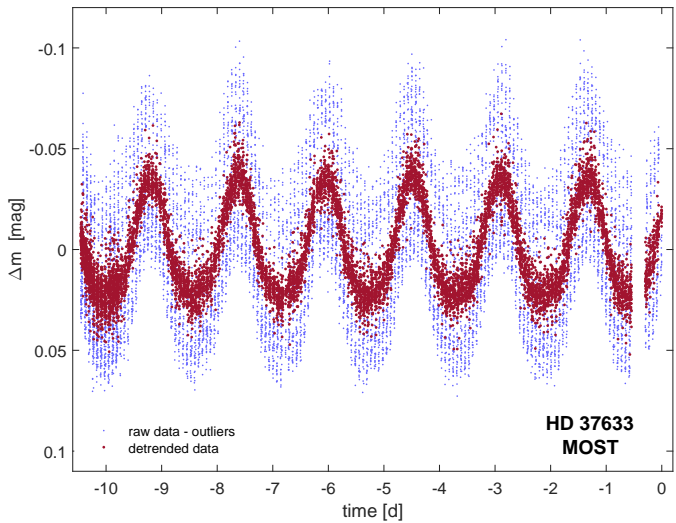}
    \caption[]{Phased light curve of HD\,37633 defined by its raw photometric data (powder-blue dots) and data detrended by our code (magenta dots). }
    \label{most5}
  \end{center}
\end{figure}
Our photometric analysis was based on 10\,572 raw MOST measurements obtained during 10.5\,d (6.7 photometric cycles). We fitted the raw measurements by a harmonic polynomial of the 5-th order using a robust regression suppressing the influence of outliers \citep[][]{mikrob}. The residuals, having an amplitude of 0.1 mag, are strongly correlated with the phase of the orbital revolution of the MOST satellite (101.48 min); nevertheless each of the 146 passages differs from each other.

Consequently, we modelled the individual course of each observational interval using linear combination of 14 special functions mentioned above (see Fig.\,\ref{most4}). Subtracting these trends, which are correlated with the MOST orbital period, we obtained detrended magnitudes defining the light curve with a scatter of 75 mmag (see Fig.\,\ref{most5}). 207~measurements were deleted. The data filtered by the MOST pipeline \citep[][]{hareter} have the same scatter as our detrended data, but the number of deleted measurements is larger: 497.

\section*{3. Conclusions}

We have shown that the proposed way of the phenomenological detrending of raw Kepler and MOST data is able to improve light curves of periodically variable stars considerably. The outlined method can be used also for the final processing of ground based (ASAS, Pan-STARRS, and SuperWASP) as well as satellite based (BRITE, CoRoT, and PLATO) data. The phenomenological detrending can be essentially applied to all variable objects whose periods and light curves are constant or regularly changing. That is valid for all rotators (CP stars, elliptical stars, spotted star, asteroids), eclipsing systems (binaries or stars transited by their planets), and several types of pulsators (RR Lyr stars, cepheids, $\delta$ Sct stars, and so on), among variable objects.

Simultaneously, we have revealed that the Kepler star \hvezda\ is definitely a common CP2 star with an effective temperature of 11\,000 K, $\log g=4.0$, and enhanced abundances of chromium, iron, and silicon, rotating with a constant period of $P=1.511\,785\,08(4)$ d. The observed light variability is then caused by the incidence of two or more photometric spots on its surface.

\section*{Acknowledgements}

The work is partly based on data collected with the KEPLER and MOST satellites. The study is supported by the grants of Ministry of Education of the Czech Republic LH14300, grant of Masaryk University MUNI/A/1110/2014, by the Czech Science Foundation grant no. GA13-10589S. This project is financed by the SoMoPro II programme (3SGA5916). The research leading to these results has acquired a financial grant from the People Programme (Marie Curie action) of the Seventh Framework Programme of EU according to the REA Grant Agreement No. 291782. The research is further co-financed by the South-Moravian Region. This work reflects only the author’s views and the European Union is not liable for any use that may be made of the information contained therein.


\begin{thebibliography}{}
\bibitem[Balona et al.(2015)]{balon}
    Balona L. A., Baran A. S., Daszy\'nska-Daszkiewicz J., De Cat P., 2015, Mon. Not. R. Astron. Soc., 451, 1445
\bibitem[Hareter et al.(2008)]{hareter}
    Hareter M., Reegen P., Kuschnig R., et al., 2008, Asteroseismology, 156, 48
\bibitem[Harmanec(1988)]{harm}
    Harmanec P., 1988, Bull. Astron. Inst. Czechosl., 39, 329
\bibitem[Jenkins et al.(2010)]{jenkins}
    Jenkins J. M., Caldwell D. A., Chandrasekaran H., et al., 2010, ApJ Letters, 713, 87
\bibitem[McNamara et al.(2012)]{mcnamara}
    McNamara B. J., Jackiewicz J., McKeever J., 2012, AJ, 143, 101
\bibitem[Mikul\'a\v{s}ek \& Gr\'af(2005)]{mik}
    Mikul\'{a}\v{s}ek Z., Gr\'{a}f T., 2005, Contr. Astron. Obs. Skalnat\'e Pleso, 35, 83
\bibitem[Mikul\'a\v{s}ek et al.(2014)]{mikcp}
    Mikul\'{a}\v{s}ek Z., Krti\v{c}ka J., Jan\'ik J., et al., in Putting A Stars into Context, Proceedings of the conference held on June 3-7, 2013, Moscow, Russia. Eds.: G.\,Mathys, E. Griffin, O. Kochukhov, R. Monier, G. Wahlgren, Moscow: Publishing house "Pero", 2014, 270
\bibitem[Mikul\'a\v{s}ek et al.(2015)]{mikbs}
    Mikul\'{a}\v{s}ek Z., Jan\'ik J., Krti\v{c}ka J., et al., 2015, ASPC 494, 189
\bibitem[Mikul\'a\v{s}ek et al.(2003)]{mikrob} Mikul\'{a}\v{s}ek Z., \v{Z}i\v{z}\v{n}ovsk\'y J., Zverko J., Polosukhina N. S., 2003, Contr. Astron. Obs. Skalnat\'e Pleso, 33, 29
\bibitem[Morris(1985)]{morris}
    Morris M. 1985, ApJ, 295, 143
\bibitem[North(1982)]{north82}
    North P., 1982, IBVS, 2208
\bibitem[North(1984)]{north84}
    North P., 1984, A\&A Suppl. Ser. 55, 259

\end{thebibliography}
\end{document}